\documentclass{vgtc}                          % final (conference style)
\ifpdf%                                % if we use pdflatex
  \pdfoutput=1\relax                   % create PDFs from pdfLaTeX
  \usepackage{graphicx}                % allow us to embed graphics files
  \DeclareGraphicsExtensions{.pdf,.png,.jpg,.jpeg} % for pdflatex we expect .pdf, .png, or .jpg files
\else%                                 % else we use pure latex
  \usepackage{graphicx}                % allow us to embed graphics files
  \DeclareGraphicsExtensions{.eps}     % for pure latex we expect eps files
\fi%

\graphicspath{{figures/}{pictures/}{images/}{./}} % where to search for the images

\usepackage{microtype}                 % use micro-typography (slightly more compact, better to read)
\PassOptionsToPackage{warn}{textcomp}  % to address font issues with \textrightarrow
\usepackage{textcomp}                  % use better special symbols
\usepackage{mathptmx}                  % use matching math font
\usepackage{times}                     % we use Times as the main font
\usepackage{cite}                      % needed to automatically sort the references
\usepackage{tabu}                      % only used for the table example
\usepackage{booktabs}                  % only used for the table example
\newcommand{\etal}{et al.}

\usepackage[resetlabels]{multibib}
\newcites{SM}{ }

\usepackage{soul}

\makeatletter
 \def\SOUL@hlpreamble{%
 \setul{}{2.2ex}%         !!!change this value!!! default is 2.5ex
 \let\SOUL@stcolor\SOUL@hlcolor
 \SOUL@stpreamble
 }
\makeatother

\onlineid{0}

\vgtccategory{Research}

\vgtcinsertpkg

\title{Data-First Visualization Design Studies}

\author{Michael Oppermann \thanks{e-mail: opperman@cs.ubc.ca} 
\and Tamara Munzner \thanks{e-mail: tmm@cs.ubc.ca}}
\affiliation{\scriptsize University of British Columbia}

\teaser{
  \centering
  \includegraphics[width=\linewidth]{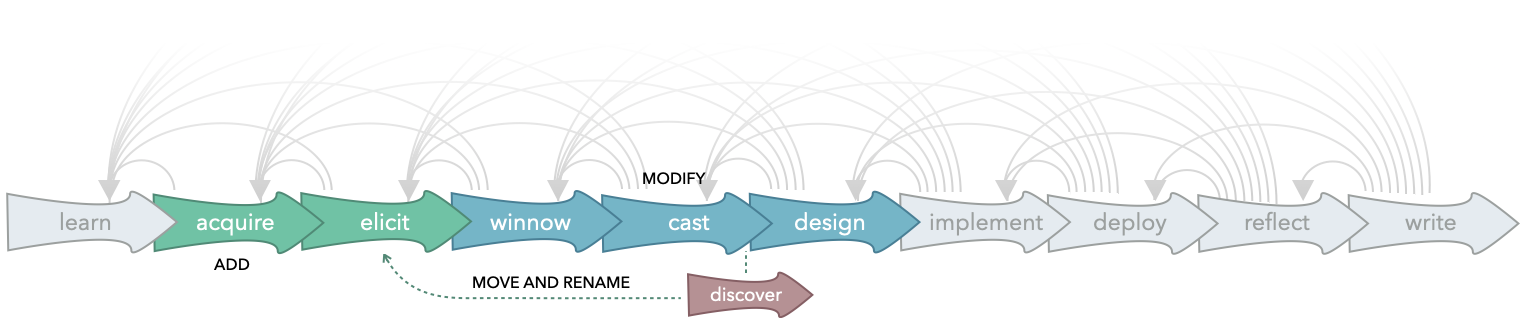}
  \caption{Refined and extended framework for data-first design studies. A new \textit{acquire} stage is added for both obtaining and abstracting data. \textit{Discover} is moved and renamed to \textit{elicit}, to emphasize the elicitation of tasks from potential stakeholders. \textit{Winnow} focuses on analyzing the match between the abstract tasks of these stakeholders and the data abstraction. Nuances differ in the \textit{cast} and \textit{design} stages to incorporate the specific characteristics of the data-first approach.}
  \label{fig:data-first-stages}
}

\abstract{We introduce the notion of a \textit{data-first design study} which is triggered by the acquisition of real-world data instead of specific stakeholder analysis questions. We propose an adaptation of the design study methodology framework to provide practical guidance and to aid transferability to other data-first design processes. We discuss opportunities and risks by reflecting on two of our own data-first design studies. We review 64 previous design studies and identify 16 of them as edge cases with characteristics that may indicate a data-first design process in action.} % end of abstract

\CCScatlist{
  \CCScatTwelve{Human-centered computing}{Visu\-al\-iza\-tion}{Visualization design and evaluation methods}{Visualization theory, concepts and paradigms}
}

\begin{document}


\firstsection{Introduction}

\maketitle

Design studies are frequently used to conduct and report problem-driven visualization research with domain experts. A design study is characterized by its highly iterative nature with a tight interplay between the analysis and abstraction of stakeholder needs, the transformation of domain-specific into domain-agnostic data, and the visualization design.

The design study methodology (DSM) by Sedlmair \etal~\cite{sedlmair2012dsm} 
provides methodological guidance on how to conduct this type of applied visualization research, proposing a nine-stage sequence: \textit{learn}, \textit{winnow}, \textit{cast}, \textit{discover}, \textit{design}, \textit{implement}, \textit{deploy}, \textit{reflect}, and \textit{write}.

The DSM focuses on connecting with collaborators in the form of domain-expert stakeholders at an early \textit{winnow} stage with suggested criteria for narrowing down the set of potential collaborators, identifying their roles in the \textit{cast} stage, and then undertaking a \textit{discover} stage focused on problem characterization and task abstraction with the chosen stakeholder. The selection of an appropriate stakeholder dictates the relevant data, which is then abstracted from domain-specific to domain-agnostic form in the \textit{design} stage, in parallel with iteratively creating appropriate visual idioms. We now characterize this DSM approach as a \textbf{stakeholder-first} ordering.

We find that this framework falls short of capturing the nature of design studies that are primarily initiated by acquiring an interesting real-world dataset rather than selecting a specific stakeholder; we call these \textbf{data-first} design studies. In a data-first design study, the early selection of the data constrains appropriate choices for stakeholders. Data abstraction is carried out early, and followed by problem characterization and abstraction for multiple potential stakeholders. Stakeholders are chosen based on whether their tasks can be supported by the selected data source.

These non-traditional data-first design studies have not been explicitly reported or analyzed in the visualization literature. However, in retrospect we realize that they occur frequently in different variations, especially in class projects or visualization design competitions~\cite{gfdrr2015vizRisk,kriebel2018makeovermonday,dataIsBeautifulDataVizBattle} but also in research contexts such as our own. We suspect that every visualization researcher has experienced the situation of stumbling across a dataset that aroused curiosity. In some cases, these serendipitous discoveries lead to the exploration of interesting design alternatives where the results and methods are shared on social media or documented in blog posts. In other cases, especially with more complex datasets or involving unique data characteristics, this process can lead to novel visualization contributions~\cite{sedlmair2016dsContributions}.

In this paper, we reflect on two of our own design studies that we now consider to be examples of a data-first approach, the \textit{Bike Sharing Atlas}~\cite{oppermann2018bikeSharingAtlas} and \textit{Ocupado}~\cite{oppermann2020ocupado}, that guided our initial thinking about this alternative methodological approach for design studies. We follow up with a review of 64 previous design studies where we note 16 edge cases that may implicitly indicate a data-first ordering. The full list of 64 studies is provided in the supplemental materials, and the 16 identified examples are provided in Appendix A. 
%but an exact categorization is not possible, as described in~\autoref{sec:literature-review}.

We contribute a first characterization of data-first design studies, namely those triggered by data instead of specific stakeholder questions. We propose an adapted version of the nine-stage design study methodology framework where we introduce a new \textit{acquire} stage, move and rename \textit{discover} to \textit{elicit}, and adjust other stages accordingly. In addition, we present a review of previous design studies and discuss opportunities and risks of these non-traditional design studies. Our goal is to provide practical guidance and aid transferability of data-first design processes.

\section{Related Work}

We discuss the extensive previous work on design studies and visualization process models to contextualize our proposed data-first approach. Literature on design studies emphasizes the need for real-world data and an early data understanding, but a stakeholder-first ordering is generally presumed.

In the original DSM~\cite{sedlmair2012dsm} the order and timing of the data acquisition is not addressed head on; the closest guidance are the questions \textit{``Does real data exist, is it enough, and can I have it?"} related to the pitfall of \textit{``no real data available (yet)"} that is noted in the winnowing stage where potential stakeholders are being assessed.

The design study \textit{lite} methodology~\cite{syeda2020designStudyLite} is an expedited version to fit within the time frame of undergraduate and graduate visualization courses. Recruiting community partners is a precondition in this framework. Although highly desirable, access to many stakeholders with course-relevant visualization needs is not always feasible. In larger class settings, students may not have access to stakeholders but could begin with data. The data-first methodology would provide a path for some or all such student groups to engage with domain experts at a later stage.

%The five design-sheet (FdS) methodology~\cite{roberts2016sketching}, commonly used in visualization courses and workshops, has been introduced to foster divergent thinking in an early ideation process. FdS requires the existence of data and can be used as a tool within a design study process to create a set of alternative designs, but in contrast with data-first design studies does not explicitly discuss the process of engaging with stakeholders. 

Munzner's nested model for visualization design~\cite{munzner2009nestedModel} and different types of process models assume a user with a specific domain problem at the start, who may act as a data provider as well. The activity-centered framework~\cite{marai2017activityCentered} suggests to mandatorily request access to real data, or a data sample of sufficient size, early in the process. McKenna et al.~\cite{mckenna2014designActivityFramework} proposed a framework that links actions directly to the nested model with the four activities, \textit{understand}, \textit{ideate}, \textit{make}, and \textit{deploy}, whereby a project can start with any activity. The \textit{understand} activity implies the data acquisition and task elicitation of the proposed data-first approach.

McCurdy et al.~\cite{mccurdy2016adr} suggest that a real-world domain problem is either expressed by domain experts or discovered by design researchers, whereas the latter is closely related to our framework adaptation. The human-centered design process by Lloyd and Dykes~\cite{lloyd2011humanCenteredGeovisualization} highlights the effectiveness of real and interesting data to engage participants in requirement elicitation, which is a key objective in data-first design studies.

Crisan and Munzner~\cite{crisan2019uncoveringDataLandscapes} proposed a conceptual framework of data reconnaissance and task wrangling to explore unfamiliar data landscapes. This framework puts data first but focuses on domain experts and how they refine data to support a specific analysis goal.

Goodwin et al.~\cite{goodwin2013creativeUserCentered} presented a creative design case study in the energy domain that explores possible visualization usage scenarios for a smart meter dataset, which is closely related to our suggested workflow but discusses primarily creativity techniques and assumes a set of users at the beginning.

In the context of evaluating visualization tools in large companies, Sedlmair et al.~\cite{sedlmair2011largeCompanies} distinguish between \textit{employee-pull} and \textit{researcher-push} solutions. Although the fundamental concept of pushing ideas from an outside perspective is similar, these authors regard it as advertising a specific tool. We instead propose to advertise data and the use of visualizations to solve a domain problem together.

\section{Case Studies}

We introduce two of our own previously published design studies
that we now consider to be examples of data-first design studies, before we generalize the underlying method in a refined DSM framework in~\autoref{sec:refined-framework} and discuss its implications in~\autoref{sec:reflections}.

The Bike Sharing Atlas combines and visualizes distributed data from several hundred bike-sharing networks worldwide~\cite{froehlich2009sensing}. We show that the data produced by these networks reveal interesting insights, not only into patterns of bicycle usage, but also the underlying spatiotemporal dynamics of a city. By working with users from different domains including a bike sharing operator, a city planner, and urban sociologists, we illustrated how interactive visualization can help to open the data that is produced in our cities to a wider audience.

Ocupado is a set of visual decision-support tools centered around occupancy data for stakeholders in facilities management and planning~\cite{oppermann2020ocupado}. We take WiFi device counts as a proxy for human presence, showing how to leverage data that had previously only been used for automation in building control systems in many new ways. We interviewed potential stakeholders from different domains and evaluated the conformity between tasks and data affordances. In a highly iterative process and with extensive feedback from a set of core stakeholders, we developed and deployed Ocupado, which was eventually adopted by our industry partner.

These two projects were different than regular design studies in the way that we engaged with stakeholders after acquiring the data.

We had worked with a bike sharing operator early on to analyze data from Vienna, but the discovery and collection of data from many station-based bike sharing networks worldwide~\cite{citybikes-api-2016} triggered additional use cases and aroused interest from new stakeholders working in different domains.
The Ocupado project began through a collaboration with a startup~\cite{sbs2020} that estimates occupancy based on WiFi signals for building automation usage. We conjectured that with suitable visualizations, this data could be actively useful to many additional user groups for decision-making and resource management, and explicitly sought out these potential stakeholders.

In both design studies we had to consider whether to embed additional data sources into the mix to support potential stakeholder tasks. For example, for the bike sharing project, we included elevation profiles because the elevation differences between stations are an essential factor for the functioning of a bike sharing system, and the combined data can provide relevant insights. When we worked with custodial managers during the Ocupado project, we were asked to add work schedules that can be compared to occupancy statistics. We ultimately decided against an integration because these schedules were not relevant for other stakeholders and it was unclear if and how they facilitate core tasks related to occupancy analysis.

\section{Refined and Extended Framework}\label{sec:refined-framework}

\begin{figure*}[ht]
 \centering
 \includegraphics[width=\linewidth]{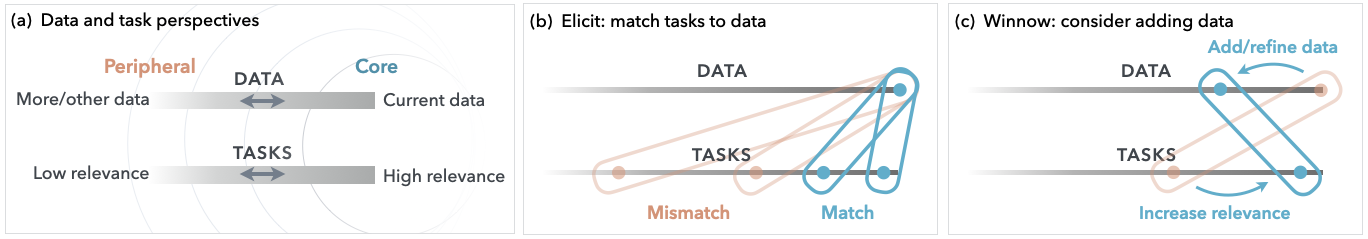}
 \caption{Simple conceptual model illustrating the challenges of finding an intersection of relevant tasks and data while keeping the problem space constrained. (a) The data and task axes both range from peripheral to core. (b) In the \textit{elicit} stage, the visualizer examines the match between the initially acquired abstracted data and stakeholder tasks. (c) In the \textit{winnow} stage, the visualizer can assess the benefits and risks of expanding the scope to additional data to expand the set of tasks and thus stakeholders.}
 \label{fig:data-and-task-centrality}
\end{figure*}

We propose an alternative, extended version of the DSM framework, generalizing our experience as a guide for future data-first design studies. 
The data-first design study stages are shown in \autoref{fig:data-first-stages}.
One major change is to add an early \textit{acquire} stage focused on both obtaining and abstracting data. The second change is to move the \textit{discover} stage immediately afterwards and rename it \textit{elicit} to reflect its task-based emphasis: obtaining information about and abstracting tasks. We keep the same ordering and general function of the \textit{winnow} and \textit{cast} stages, though \textit{winnow} is heavily dependent on the initially acquired data, and nuances differ in the \textit{cast} stage. The focus of the \textit{design} stage is narrowed to cover only idioms, since the data abstraction occurs earlier in the new \textit{acquire} phase. The subsequent stages \textit{implement} and \textit{deploy} are unchanged, as is the initial \textit{learn} stage. \textit{Reflect} and \textit{write} are adjusted to reflect and report also on the peculiarities of the data-first process. As with the original DSM framework, jumping backwards to previous stages is common and necessary, in particular when adding new stakeholders. 

We use the term \textit{visualizer} to mean either a visualization researcher or practitioner. The adapted framework contains 10 stages:

\vskip 5pt
\noindent \textbf{Learn:} This stage remains unchanged from the original because it concerns the visualization knowledge in general and independent of the specific domain problem. A solid understanding of the visualization literature is crucial for every design study.

\vskip 5pt
\noindent \textbf{Acquire data:} A visualizer encounters, collects, generates, or obtains access to a dataset. The visualizer translates the data description into domain-independent language and begins to develop a data abstraction. Data sketches and descriptive statistics can provide first insights and guide the development of initial hypotheses about its underlying semantics.

The data-first DSM framework emphasizes iteration, as does the original DSM, and thus takes into account the idea that further data abstraction refinement will surely occur after task abstractions are identified. Relevant questions: \textit{What type of data am I working with and what are the underlying semantics? Are there any data quality challenges and is long-term access guaranteed? What is special about this data and who would benefit of seeing and exploring it?}

\vskip 5pt
\noindent \textbf{Elicit tasks:} The visualizer seeks out multiple potential stakeholders and elicits domain-specific tasks from them that might be relevant for the chosen data abstraction.  The visualizer both explains the initial data abstraction and learns about unsolved stakeholder needs.

The visualizer translates the tasks into domain-independent language and considers what data is required to address them. \textit{Is there a match between given data and stakeholder need? Is more or other data required to answer domain-specific questions, and if so, how difficult is it to acquire and integrate?}

\vskip 5pt
\noindent \textbf{Winnow:} The goal of the data-first DSM winnowing stage is to assess and prioritize the set of potential stakeholders according to the match between their abstract tasks and the data abstraction. This process may involve multiple rounds of task and data abstraction, as the understanding of the stakeholder needs and of the dataset semantics evolves from partial to more complete. We discuss these data and task perspectives in~\autoref{sec:winnow-tasks-refine-data}.
\textit{How frequent are their data-relevant tasks? How central are these tasks to the stakeholder's primary mission? How many people in the organization deal with these tasks?}

\vskip 5pt
\noindent \textbf{Cast:} The \textit{cast} stage is similar to the original DSM with the goal to identify collaborator roles; here we note some previously proposed roles that have particular ramifications in the data-first case. 

We named the \textit{promoter} role in the Ocupado paper~\cite{oppermann2020ocupado}, and call it out here as essential in a data-first design study. The visualizer needs to serve in this role to reach out to potential stakeholders to present data and visualization opportunities. In addition, partners or stakeholders may act as promoters themselves when they undertake technical evangelism to other potential users about the benefits of a visualization system. When this evangelism includes demos of software prototypes, as it often does, the promoter may be involved as an intermediary in the task elicitation and winnowing stages. In this case, the promoters can serve as a conduit to relay information back to the visualizers about stakeholder needs and how they align with prototype capabilities. In retrospect, we identify an example of this role occurring within our own previous Overview work~\cite{brehmer2014overview}.

The \textit{data producer} and \textit{data consumer} roles were first proposed by Kerzner et al.~\cite{kerzner2015vulnerabilityAnalysis} within  stakeholder-first design studies. These two roles are most certainly held by different people in data-first studies, and therefore require particular consideration. A data provider may need to ensure long-term access to and the functioning of the data pipeline. The visualizer may sometimes take on the data provider role in a later project stage, even if that role was played by somebody else at an early stage. 

\vskip 5pt
\noindent \textbf{Design:} The \textit{design} stage focuses only on visual idioms, whereas it also includes data abstraction work in the original DSM; in our framework, the latter was carried out earlier, in the \textit{acquire} stage. 

\vskip 5pt
\noindent \textbf{Implement} and \textbf{deploy:} Remain unchanged from the original DSM.

\vskip 5pt
\noindent \textbf{Reflect} and \textbf{write:} Data-first design studies invite specific considerations during the reflection and writing stages. 
%specific layers of reflection to foster transferability, to ensure rigor~\cite{meyer2019criteriaForRigor}, and to create a meaningful impact for the visualization community. 
\textit{How many potential stakeholders and domains have been considered, and ultimately chosen? What are the differences and commonalities among the stakeholder tasks? Did the original data suffice or was there a need to integrate data from additional sources?}

\section{Reflections and recommendations}\label{sec:reflections}

We reflect on the challenges of the winnowing stage, the opportunities of the data-first approach, and its risks. 
%In the following section, w
%We discuss implications of data-first design studies by reflecting on our own work.

\subsection{Winnowing Tasks and Refining Data}\label{sec:winnow-tasks-refine-data}

The key challenge in this non-traditional design study approach is to identify which stakeholders are appropriate, and thus to identify whether there is a match between potential stakeholder tasks and the available data. To do so, the visualizer must assess the relevance of the identified task for the given data, and also consider whether to integrate secondary data sources. These considerations take place during the \textit{winnow} stage. 
%We use these perspectives as reference to discuss the balance between data and tasks.

\autoref{fig:data-and-task-centrality} illustrates a simple conceptual model for reasoning about the conformity between data and tasks as aligned axes through a continuum with peripheral on the left, and core on the right. At the core are the current data and the high-relevance tasks, with other data and low-relevance tasks being relegated to the periphery.
\autoref{fig:data-and-task-centrality}b shows several tasks assessed according to the initial data selection arising from the \textit{elicit} stage, with some determined to be mismatches and some considered as matches. 

~\autoref{fig:data-and-task-centrality}c illustrates the tradeoffs of incorporating additional datasets into the scope of a project during the \textit{winnow} stage. A possible benefit of adding secondary data sources could be to support more tasks, and thus to expand the set of eligible stakeholders. However, the visualizer also needs to ensure that the problem space under consideration maintains sufficient cohesion to have clear system goals and design targets. Cohesion can be reasoned about as constraints on the number and variety of stakeholders and of tasks.

%In this conceptual model, we deliberately omit the visual encoding and concentrate on the data and task perspectives because their linkage is the decisive factor when selecting stakeholders. Nevertheless, there are important implications for visual encodings and the system design in general. An adaptable and extensible visualization facilitates data iteration and potentially a better support of tasks. \tm{i don't understand last two sentences. maybe kill whole paragraph??}

\subsection{Opportunities}\label{sec:opportunities}

We have identified several opportunities of a data-first approach. 

\vskip 5pt
\noindent \textbf{Capitalizes on possibilities that do not fit traditional DSM process.} We consider data-first design studies as an alternative way to approach visualization research collaborations, where interesting research questions can emerge through conversations that unfold later in this process than in the traditional one. Analogous to market pull versus technology push dynamics in product innovation strategies~\cite{baxter1995productDesign}, the originally acquired data is used to push ideas from an outside perspective instead of pulling analysis questions from a domain expert. For example, using existing WiFi access points to sense occupancy patterns in hundreds of rooms without installing additional hardware yielded a completely new opportunity that Ocupado stakeholders had not envisioned before.

%Although there is a risk of talking to many stakeholders without identifying compatible tasks, they may have a need that implies an interesting visualization research question and the conversations will open the door for future collaborations.

Launching a data-first design study without a specific stakeholder collaboration commitment at the beginning of a project is a possible method to approach long-distance work and to engage with domain experts in imperfect circumstances, such as under COVID-19 restrictions. The \textit{learn} and \textit{acquire} stages are conducted independently by the visualizer. When reaching out to potential stakeholders, first ideas for solving a specific problem are presented and guide the discussion, instead of conducting a fully open-ended brainstorming session which can be challenging using only digital communication tools.

\vskip 5pt
\noindent \textbf{Allows very early data sketches or technology probes with real-world data.} Launching a design study project without access to real data is a common pitfall~\cite{sedlmair2012dsm}. Synthetic or toy data often results in additional project obstacles and delays~\cite{kerzner2015vulnerabilityAnalysis, williams2019visualizingMovingTarget}, and may even lead down the wrong path of data abstraction and visualization design. Access to real-world datasets at the beginning of the project allows the creation of data sketches~\cite{lloyd2011humanCenteredGeovisualization} and technology probes~\cite{hutchinson2003technology}, both to internally guide the development of hypotheses and possible usage scenarios, and to externally inquire needs and desires of target user groups.

\vskip 5pt
\noindent \textbf{Supports gradual expansion of stakeholder set.} Multi-channel engagement with several stakeholders fosters serendipity and enables more impactful visualization designs, as demonstrated by Wood et al.~\cite{wood2014moving} and confirmed in our own data-first design studies~\cite{oppermann2020ocupado, oppermann2018bikeSharingAtlas}, while providing breadth to a convincing evaluation~\cite{sedlmair2016dsContributions}. However, addressing the needs of multiple stakeholders involves the risk of conflicting tasks and an ever-expanding project scope. In data-first design studies, stakeholders are selected by evaluating the conformity between tasks and data capabilities. Although additional data could be added, the primary data source serves as a guide to navigate the due diligence with stakeholders and to delimit the problem space.

\vskip 5pt
\noindent \textbf{Encourages progression of course or side projects into publishable results.} Data-first design studies may start off during visualization courses or when exploring and visualizing data out of curiosity. Characteristics such as complex datasets, unique characteristics, or combinations of different data sources that address high-relevance stakeholder tasks may suggest a fruitful direction to expand initial efforts to result in interesting and novel knowledge contributions.

\subsection{Risks}\label{sec:risks}

\vskip 5pt
\noindent \textbf{Hypothesized tasks.} In the data-first approach, visualizers begin exploration immediately after acquiring the initial data, with conjectures about potential stakeholders and usage scenarios. We consider the delayed involvement of potential domain experts to be a major risk, because the initial implementation is based on hypothesized rather than verified tasks. To mitigate this risk,  %Autobiographical design will not evolve into a user-centered design study, as discussed in~\autoref{sec:discussion}. 
technology probes should be limited in functionality and open-ended with respect to use to avoid wasted effort~\cite{hutchinson2003technology}.

\vskip 5pt
\noindent \textbf{Hammer looking for a nail.}

The danger of having data at our fingertips is to converge on a specific idea prematurely and develop a full-fledged prototype without an in-depth engagement of stakeholders and the consideration of the broader design space. In a design study, stakeholder participation needs to be maximized and many design alternatives must be considered, otherwise stakeholders are marginalized to design verifiers.

\vskip 5pt
\noindent \textbf{No actual users.} In the worst case, the search for potential stakeholders may not succeed. In this case, other types of research contributions such as a novel visualization technique or algorithm may be possible, but a successful design study would not be achievable because it requires working with real users to solve their real-world problems. A major pitfall would be attempting to validate design decisions prematurely based on speculative tasks; care should be taken to avoid this mistake. A less extreme situation is when significant time is spent on promotion, leaving less energy available to direct towards research questions. The pitfall in this case would be a design study that requires longer than planned before the work is sufficiently complete to publish. 

\vskip 5pt
\noindent \textbf{Mismatch between data and stakeholder tasks.} A clear match between data and stakeholder tasks is not always immediately obvious and a short-cut to data and task abstractions can lead to a premature commitment to specific stakeholders. In some cases, proxy measures that stand in for a variable of interest entail a range of limitations. In other cases, the given data is out of date or needs to be augmented with other data sources. A misinterpretation can lead to project detours or to a failed design study.

\vskip 5pt
\noindent \textbf{Joy of discovery versus actual needs.} The presentation of data through visualizations often sparks interest, particularly when potential stakeholders see their own data or data similar to theirs. We observed many meetings where domain experts enthusiastically began to conjecture about the many different ways they might be able to use the data. This process is encouraged to assess the conformity between the given data and relevant tasks. However, at these initial meetings ideas are often proposed based on limited information; the challenge for the researcher is to discriminate between initial enthusiasm and substantive user needs. A central concern is to understand whether the visualization addresses a core task of the target user group, to avoid the pitfall of a task that is only of peripheral relevance. Fellow tool builders and gatekeepers~\cite{sedlmair2012dsm}, who frequently act as intermediaries in \textit{winnow}-stage meetings, might misunderstand specific needs of front-line analysts. The pitfall to avoid is premature selection of stakeholders whose actual tasks do not in fact align with the data, who would then abandon a visualization solution quickly.

\vskip 5pt
\noindent \textbf{Data promises.} Decoupling the data producer from the consumer may increase the responsibility for the visualizer who needs to ensure the functioning of the data pipeline throughout all stages of the design study---as opposed to stakeholder-first design studies where data is typically provided by or collected together with the domain expert. A risk is therefore to have a strong collaboration with stakeholders but not deliver on the initial data promises. For example, generating interest by presenting data from one city, but not being able to follow through by collecting data from the city that is most relevant to the stakeholder, could be a major barrier for project success that is independent of the quality of the proposed visualization solution.

Although in many cases setting up a static database at the beginning is sufficient, for projects that rely on the availability of continuous data streams, external API or system level changes can jeopardize the success of a design study. To guarantee long-term availability and avoid continuous engineering effort, it is imperative to offload this responsibility.

Stakeholders likely depend on data providers even when the research project concludes. Visualization researchers may, but do not necessarily, act as data providers in a data-first design study. 

\vskip 5pt
\noindent \textbf{Closed design approach.} Data iteration~\cite{hohman2020dataIteration}, including the integration of secondary data sources in a visualization system, can be a crucial step to address a specific domain problem. One major risk in data-first design studies is to keep the focus too narrow on the initial data without considering the wider problem space and ensuring extensibility and adaptability from an engineering and design perspective.

\section{Review of Previous Design Studies}\label{sec:literature-review}

In parallel with formalizing our proposal for a refined data-first design study methodology framework, we conducted a literature review of a diverse set of design studies. While we had our own direct experience of two projects that started with a dataset rather than a specific stakeholder as an existence proof, the ratio of these studies in the previous visualization literature was unclear. We were interested whether data-first approaches have been explicitly or implicitly described previously, what type of data they have used, and if there are any underlying common themes.

In total, we selected 64 design studies that were published between 1999 and 2019 (full list provided in supplemental material). We used the work by Lam et al.~\cite{lam2017goals2Tasks} who identified 39 design studies as a seed and extended this list with additional examples from Sedlmair~\cite{sedlmair2016dsContributions} and by reviewing previous conference proceedings. In addition, we sourced design studies from \mbox{vispubdata}~\cite{isenberg2016vispubdata} by searching for the term ``design study" in titles and abstracts. The vast majority of these design studies are published at InfoVis with a few examples from VAST and short papers.

In our analysis we identified 16 edge cases (25\%) that indicate characteristics of a data-first design process in the motivation or process statements, listed in Appendix A. An accurate allocation to a stakeholder-first or data-first ordering is not possible based on this kind of retrospective interpretation. Nevertheless, these edge cases provide relevant insights and illustrate that data-first approaches are not at all unusual, although they are often underdocumented.

These potential instances of data-first design studies span a broad range of topics and data sources, such as music listening histories~\cite{baur2010streamsOfOurLives}, code repositories~\cite{ogawa2009code_swarm}, electricity consumption~\cite{van1999cluster, goodwin2013creativeUserCentered}, financial transaction flows~\cite{arleo2019sabrina}, sports~\cite{perin2013soccerstories, polk2014tennivis}, social media~\cite{miranda2016urban}, hotel customer reviews~\cite{wu2010opinionseer}, and genealogies~\cite{bezerianos2010geneaquilts}. The data is mostly acquired from publicly available databases but in some cases scraped or generated by researchers.

For instance, Miranda et al.~\cite{miranda2016urban} collected Flickr activity and tweets to capture spatio-temporal patterns in a city. The authors illustrated how interactive visualizations of this data can help experts specializing in different domains to better understand people’s behavior in public places.

The BallotMaps~\cite{wood2011ballotmaps} design study used electoral data from an open data repository to identify name and ordering biases in ballot papers. External domain experts are not apparent based on the publication but the study demonstrates with real-world examples how a visual approach can reveal significant biases that were not evident from statistical analysis alone.

Distinguishing between researchers being their own stakeholders and data-first design studies is most difficult, as there is sometimes no clear separation. Researchers with extensive expertise in a specific domain problem may be qualified to self-validate a visualization solution in an \textit{autobiographical} design study, as we discuss in~\autoref{sec:autobiographical-design-studies}. In these 16 edge cases it is often unclear if the researchers stumbled upon data because of a broad interest in a topic or if they had a particular analysis question themselves. The level of domain expertise is sometimes indicated, for example, one of the authors of TenniVis~\cite{polk2014tennivis} had 35 years of tennis playing experience, but in most design studies it is not evident.

While some authors~\cite{perin2013soccerstories, wu2010opinionseer, basole2013interfirmRelationships} report on early interviews with domain experts to elicit requirements, many design study projects include external stakeholders rather late in the process, for example, to participate in a summative evaluation instead of influencing the design process. We consider this late involvement of target users a major risk (see ~\autoref{sec:risks}), particularly for data-first approaches when real data is available and simple data sketches progress to more complex solutions.

While this review is an attempt to analyze many different types of design studies from the last two decades, our selection is certainly incomplete as many other design studies have been published. A further limitation is that our judgement is based on limited information on the design process and additional investigations are required to better understand implications of data-first approaches that were not explicitly reported. Interviewing visualization researchers would be a great opportunity for future work.

\section{Discussion}\label{sec:discussion}

We discuss the role of the researcher's domain expertise, and design studies without engaging in collaboration with external domain experts.  

\subsection{Domain Familiarity and Data Understandability}

The nature of data-first design studies requires a certain degree of domain familiarity and data understandability. Thus, not all datasets are appropriate for data-first design studies: the selection strongly depends on the personal background of the visualizer as it pertains to the research context. Initial data should be self-explanatory or well-documented, or the visualizer should have a strong personal interest.

Visualizers acquire and investigate data on their own before engaging into conversations with potential stakeholders. While specific use cases might not be immediately apparent and talks with different domain experts can lead to serendipitous findings, visualizers need to be aware of problems faced by domain experts and foreshadow potential tasks at the beginning of a project. We argue that visualizers need to be \textit{task-curious} to advance this type of problem-driven research, which is methodologically very different to traditional design studies where only stakeholders come forward with tasks.

A thorough understanding of the selected data and its limitations is essential. Although this understanding will typically grow over time, a basic knowledge at the beginning is beneficial. Descriptive statistics and superficial explorations quickly reach their limits. In both our data-first design studies, we recorded data based on a live data stream. In the bike sharing project, although we recorded data from hundreds of cities, it followed a relatively simple and uniform scheme which allowed us to detect data limitations without having a deep domain knowledge. In the Ocupado project, we worked closely with our industry partner and data provider to better understand data semantics and quality issues. 

\subsection{Data Repurposing}\label{sec:data-repurposing}

The data-first approach entices visualizers to use available data and apply it to new problem contexts. For example, proxy variables may be used to answer questions when  direct measurement is prohibitively expensive \cite{montgomery2000measuringLivingStandards}. The repurposing of data~\cite{woodall2017dataRepurposing} provides unique opportunities but entails a range of risks: data collectors may not have given consent for their data to be used in a particular way, the original data collection process may not be fully transparent and thus the data quality and underlying assumptions can not be judged reliably, and a lack of correlation with the variable of interest can ultimately entice analysts to draw wrong conclusions. Visualization researchers are responsible to determine if data is fit for the new use and need to understand all its ethical and analytical implications. As discussed in ~\autoref{sec:winnow-tasks-refine-data}, one of the goals in the winnowing phase is to identify if the data is appropriate for a given task or if it would be better served with a different dataset.

\subsection{Autobiographical Design Studies}\label{sec:autobiographical-design-studies}

Following the terminology of Neustaedter et al.~\cite{neustaedter2012autobiographical}, we define the process of designing and evaluating visualizations through self-usage as \textit{autobiographical} design studies, that we distinguish from \textit{data-first} design studies.

Although the DSM paper does not use this terminology, which was introduced at roughly the same time as its publication, the idea of it is clearly articulated within it: ``While strong problem-driven work can result from situations where
the same person holds both of these roles [visualization researchers and domain experts], we do not address this case
further here"~\cite{sedlmair2012dsm}.

Autobiographical design draws on ``extensive, genuine usage by those creating or building the system"~\cite{neustaedter2012autobiographical}. Visualization experts act as primary stakeholders and design iterations are based on their own experiences. This type of first-person research~\cite{desjardins2018revealingTensions} usually emerges from an unmet personal need and allows tinkering with an idea in a research context. The researcher should have or quickly gain extensive expertise related to the domain problem in order to validate the effectiveness of a visualization.

In contrast, data-first design studies follow the autobiographical design only at the very beginning. Domain familiarity is important but the process largely focuses on external domain experts~\cite{wong2018domainExperts} as target users and their feedback shapes the design decisions instead of insights from self-usage. 

\section{Conclusion}

We contribute the first characterization of a data-first design study for applied visualization projects that begin with an interesting real-world dataset rather than stakeholder selection. The motivation behind this approach stems from our own experience in conducting two data-first design studies.
Problem-driven research motivated and inspired by data enables opportunities that do not fit in the traditional design study process, although it does introduces risks and tensions. 
We propose an adaptation of the design study methodology framework to provide practical guidance for researchers interested in this alternative way to approach research collaborations.
A preliminary review of published design studies revealed that roughly one quarter of them were edge cases which might have had an implicit data-first ordering. We conjecture that some of these researchers may have under-emphasized their true trajectory by bending their explanation closer to the previously articulated design study methodology framework. We hope this work encourages researchers to be more explicit in documenting and reflecting on various flavors of design studies.

\acknowledgments{
We thank J\"urgen Bernard, Madison Elliott, Steve Kasica, Zipeng Liu, Francis Nguyen, and Ben Shneiderman for inspiring discussions and feedback. We also thank the anonymous reviewers for their comments.}

\bibliographystyle{abbrv-doi}

\bibliography{template}

\appendix
\section{Appendix: Edge Case Design Studies}

\vspace*{-15pt}
\bibliographystyleSM{abbrv-doi}
\nociteSM{*}
\bibliographySM{appendix}

%\nobibliography*
%\bibentry{a-polk2014tennivis}.
%\bibliography{appendix.bi}

%\bibliographystyle{plain}
%\nocite{*}
%\bibliography{appendix}

\end{document}